\documentstyle[epsf,aaspp4]{article}
\begin{document}

\lefthead{K. M. Sealey et al.}
\righthead{Observations of QSO J2233-606 in the Southern Hubble Deep Field }

\title{Observations of QSO J2233-606 in the Southern Hubble Deep Field}
\author{K. M. Sealey,
M. J. Drinkwater,
and J. K. Webb}
\affil{Department of Astrophysics and Optics, School of Physics,
University of New South Wales, Sydney 2052, Australia}

\begin{abstract}
The Hubble Deep Field South (HDF-S) Hubble Space Telescope (HST)
observations are expected to begin in October 1998.  We present a
composite spectrum of the QSO in the HDF-S field covering
UV/optical/near IR wavelengths, obtained by combining data from the
Australian National University 2.3m Telescope with STIS on the Hubble
Space Telescope (HST)\footnote{Our ground-based 2.3m spectrum and the
composite spectrum are available in electronic format at the address
http://bat.phys.unsw.edu.au/$\sim$kms/hdfs/}.  This intermediate
resolution spectrum covers the range 1600--10000\AA~ and allows us to
derive some basic information on the intervening absorption systems
which will be important in planning future higher resolution studies
of this QSO.

The QSO J2233$-$606 coordinates are 22:33:37.6 $-$60:33:29 (J2000),
the magnitude is B=17.5, and its redshift is z$_{em} = 2.238$, derived
by simultaneously fitting several emission lines.  The spectral index
is $\alpha = -0.7 \pm 0.1$, measured between the Lyman-$\alpha$ and Mg
II emission lines.  Many absorption systems are present, including
systems with metal lines redwards of the Lyman-$\alpha$ emission line
at $z_{abs} = 2.204, 1.942, 1.870, 1.787$ and a few very strong
Lyman-$\alpha$ features at $z_{abs} = 2.077, 1.928$, without similarly
strong metal lines.  There is a conspicuous Lyman limit (LL)
absorption system that is most likely associated with the $z_{abs} =
1.942$ system with a neutral hydrogen column density of
N$_{HI}=(3.1\pm1.0)\times10^{17} cm^{-2}$.  There is some evidence
for the presence of a second LL absorber just to blue of the
conspicuous system at $z = 1.870$.  We have employed a new technique,
based on an analysis of the shape of the observed spectrum in the
region of the LL absorption, to explore the properties of the gas.  We
tentatively conclude that this system might have suitable
characteristics for measuring the deuterium-to-hydrogen (D/H) ratio.

\end{abstract}
\section{Introduction}

Compared to the original HDF, scientific studies of the the Hubble
Deep Field South (HDF-S) may benefit from the additional information
provided by a bright high redshift $z>2$ QSO lying close to (but not
within) the targeted WFPC-2 imaging area.  Initially 12 small
($1\deg~\times~1\deg$) fields were selected as candidate target
fields.  Whilst no known high $z$ QSOs were present in any of the
possible fields, Hewett and Irwin found a high $z$ QSO candidate in
one of the fields using a UK Schmidt Telescope IIIaJ objective prism
plate scanned by the Automated Plate Measuring facility in Cambridge.
Follow-up observations of the candidate at the Anglo-Australian
Telescope confirmed the object to be a QSO (J2233$-$606 22:33:37.6
$-$60:33:29 (J2000), B=17.5) with a redshift $z>2$ (see Boyle 1997 for
details).  We note for completeness that 20cm radio observations have
been made (Ray Norris, private communication reported in Boyle 1997)
that give an upper limit on the radio flux as S$_{20cm}<3$mJy.  There
is a strong radio source 10 arcmin from the QSO (S$_{20cm}=146$mJy).

We present here a low-resolution spectrum of the HDF-S QSO taken with
the Australian National University (ANU) 2.3m Telescope and combined
with the publicly available STIS spectrum providing wide spectral
coverage.  The QSO has an optically thin Lyman limit which is a
possible candidate for measuring the deuterium-to-hydrogen (D/H)
ratio. The QSO also has several strong absorption features redwards of
the Lyman-$\alpha$ emission line, to which we assign tentative
identifications.

\section{Observations and Data Reduction}

The ANU 2.3m observation of the QSO was taken on 1997 October 21 at
Siding Spring in clear conditions with $\approx$~2 arcsec seeing.  The
Double Beam Spectrograph was used with a 300~line/mm grating in the
blue arm (binned by 2 in dispersion), and a 158~line/mm grating in the
red giving a resolution of 8~\AA\ in each arm. A dichroic was used to
split the light at $\approx$~5400~\AA. Due to the difficulty in
combining the red and blue spectra around the dichroic wavelength
there is some uncertainty in our spectrum in the wavelength region
$\lambda 5400-5600$\AA. The data were reduced using standard
procedures with the IRAF \footnote{IRAF is distributed by the National
Optical Astronomy Observatories, which is operated by the Association
of Universities for Research in Astronomy, Inc. (AURA) under
cooperative agreement with the National Science Foundation.} image
analysis software.  The resulting spectrum ($\lambda 3300 -
10000$\AA) is the weighted combination of 2 exposures (1200s and
900s).  

We combined our ground-based 2.3m spectrum with the HST STIS spectrum
taken on 1997 October 30 with the G230L grating for 3700s ($\lambda
1600-3100$\AA\, resolution $\approx3.4$\AA) and the G430L grating for
2200s ($\lambda 2900-5700$\AA, resolution $\approx$6\AA).  Our 2.3m
spectrum was not spectrophotometric (we used a narrow slit) so we
scaled it before combining to have the same flux level as the STIS
spectrum in the overlapping region of the two spectra.  In order to
preserve the higher resolution of the STIS data and because of
uncertainties in the flux calibration of the ground-based data in the
blue, we only used data redward of the Lyman-$\alpha$ emission
(4170\AA) from the 2.3m data in the combined spectrum. A comparison of
the 2.3m spectrum and the STIS spectrum in the overlap region made by
cross-correlating the two spectra does not show any significant
wavelength shifts, with a typical uncertainty of 40 km s$^{-1}$
(0.5\AA).  This is consistent with the rms residuals obtained for the
comparison lamp exposures for the 2.3m blue DBS spectrum, which were
around 0.6\AA.

The final composite spectrum (shown in Fig.~1) has been rebinned to
the dispersion of the UV STIS component of the spectrum
($\lambda<2900$), 1.55\AA/pix.  The signal-to-noise per pixel is
$\approx$~5 below the Lyman-limit break ($\lambda<2700$), $\approx$~20
in the Lyman forest, and up to $\approx$~40 around $\lambda 7000 -
8000$\AA.

\placefigure{fig1}

\section{Results}

\subsection{Emission lines and QSO continuum}

We measured the redshift of each of the emission lines in the QSO
independently and also determined the QSO redshift by simultaneously
fitting some of the emission lines using VPFIT (Webb 1987, Carswell et
al.\ 1992). Table 1 shows, for each of the emission lines, the
redshifts, the equivalent widths (EW) and the velocity widths.  The
four lines used in the simultaneous measurement were C III] $+$ Al
III, C IV and Mg II.  The individual redshifts for each line are
listed in Table 1. The redshift from the simultaneous fit is z=2.238.
The difference of 1390~km s$^{-1}$ between the (C IV and C III] $+$ Al
III) and Mg II redshifts is not unusual (Espey et al.\ 1989).  The
observed EWs lie in the ranges given by the composite QSO spectra of
Francis et al.\ (1991) and Boyle (1990).

\placetable{tbl1}

The spectral index of the combined spectrum is $\alpha=-0.7\pm0.1$
(where $f_\nu\propto\nu^\alpha$), consistent with values derived for
optically selected QSOs (eg. the LBQS survey yields $0.0<\alpha<-0.8$;
Francis et al.\ 1996).  The measurement was made by fitting to
three sections of the spectrum between $\lambda$4000$-$9000\AA~ free
of detectable emission lines.  Our estimate of the uncertainty in
$\alpha$ was derived by fitting the combined spectrum and the 2.3m
data alone and hence is approximate.

\subsection{Absorption systems}

An initial investigation of the QSO reveals that there is complex
absorption present from intervening systems. The absorption system
details are tabulated in Table 2. There are four systems with metal lines
and two systems with no detectable metals 
(at the signal-to-noise and resolution of this data) redwards of the
Ly-$\alpha$ emission line.

There is a Ly-$\alpha$ line at z=1.870 (L3) with associated metal
lines (CII, SiIII and strong CIV).  If NV is present, it is blended
with the stronger (z=1.928) of the 2 Ly-$\alpha$ lines at z=1.928 (L1)
and z=1.942 (L2) (see top panel of Figure 2) which we are associating
with the LL break (discussed later).  We note that the CIV absorption
line associated with L3 (z=1.870) line falls blueward of the SiIV
emission line so a SiIV identification is also feasible.  However,
this seems unlikely because there is no corresponding Ly-$\alpha$
absorption line for a SiIV identification.

Identifications of other metal systems are less secure due to there
being fewer metal lines.  Prominent Ly-$\alpha$ absorption features at
z=1.942 (L2), z=1.787, z=2.204, all appear to have CIV absorption.
Also, there are absorption lines lying mid-point between the
Ly-$\alpha$ and NV emission lines which may be due to NV absorption
corresponding to the z=2.204 system.

There appear to be no detectable metal lines associated with the
strong Ly-$\alpha$ line (z=2.077), or with L1 (z=1.928).  The strong
line at 3003.9\AA~ is due at least in part to Ly-$\beta$ associated
with L1 (z=1.928).

If the identifications above are correct, we are seeing a rich complex
of CIV absorption spread over the redshift range 1.787 to 2.204
corresponding to a scale $\sim 100$ Mpc, if interpreted as a spatial
separation.  Detailed studies of this QSO and the region of the sky
around it, particularly photometric redshifts analyses, should reveal
whether we have discovered a huge structure formed at an early epoch,
or whether the line of sight to this QSO happens to intersect multiple
isolated galaxies or clusters.

\placetable{tbl2}

The STIS spectrum can be used to explore the properties of the Lyman
limit absorption system to determine the neutral hydrogen column
density, N$_{HI}$, and to constrain the velocity width of the
absorbing gas.  This is of particular interest from the point of view
of seeing whether the system has suitable characteristics for
measuring the deuterium to hydrogen ratio (D/H).  The important
requirement, apart from having a sufficiently high N$_{HI}$ for DI to
be detectable, is that the DI feature is not blended with an
ill-placed HI line, blue-shifted with respect to the main system.
Since the DI feature is blue-shifted with respect to HI by $-81$ km
s$^{-1}$, any absorption complex with velocity structure broader than
this may not be useful for a D/H determination (unless by chance a
high neutral hydrogen column density component happens to lie on the
blue side of the complex, relatively well separated from other
components).  To explore the extent of the velocity spread in the LL
absorber here, we introduce a method that utilizes the shape of the
spectrum at the Lyman limit to estimate the width of the complex.
Lyman limit absorption systems which have a low b-parameter and a
single high column density component will produce a sharp Lyman limit
discontinuity.  Conversely, a high b-parameter or complex absorption
will produce a more gradual decline at the Lyman limit.

We first estimate the neutral hydrogen column density (N$_{HI}$) of
the absorption system causing the depression at the Lyman limit using
N$_{HI}\sigma_{LL}=\tau$ where $\sigma_{LL}=6.3~\times~10^{-18}
cm^{2}$ and $-lnI_{o}=\tau$ ($I_{o}$ is the ratio of the flux in the
spectrum to the flux in the continuum just bluewards of the drop in
the LL (2683\AA) and $\tau$ is the optical depth).  To estimate the
approximate uncertainty in N$_{HI}$ due to errors in the placement on
the continuum, we derived N$_{HI}$ for two different continua (the
lower of these, the preferred one, is shown in Fig.~1).  The value
derived is N$_{HI}=(3.5\pm0.3)\times10^{17}cm^{-2}$.

There are additional uncertainties involved with our estimate of
N$_{HI}$; a general background of Lyman-$\alpha$ and higher order
absorption lines will also contribute to the flux decrement we measure
below 2683\AA.  At $z \sim 2$, this effect means the estimate of
N$_{HI}$ above is systematically high by approximately 10-20\%.

To identify the Lyman-$\alpha$ line associated with the Lyman limit
(LL) we used the wavelength (2685\AA) of the LL to 
estimate where the associated Ly-$\alpha$ line would fall.  Fig. 2
shows there are two candidate lines, one at z=1.928 (L1) and
the other at z=1.942 (L2).  They are sufficiently close to each other
that it is unclear which is associated with the LL.

In order to explore whether L1 or L2 is associated with the LL, we
generated a series of simulated spectra at the column density derived
above, for a range of velocity dispersion parameters, (b = 10, 30 and
60 km s$^{-1}$, where b = $\sqrt(2)\sigma$) at the two redshifts of L1
and L2.  We then compared the agreement between the model and observed
spectra in the LL region.  The low spectral resolution makes it
difficult to derive useful information by comparing model and observed
individual Lyman series lines.  Instead, a more promising method is to
look explicitly at the wavelength and shape of the LL drop, at and
around the point at which the flux falls to roughly a constant level.
The second and third panels of Figure 2 show the 2 possible 
systems associated with the LL with the models overlaid.

\placefigure{fig2}

Fig. 2 shows that it is still difficult to unambiguously determine
whether the actual Lyman limit absorber is associated with L1 or L2.
However, if it is associated with L2, the b=30 and 60 km s$^{-1}$
models are inconsistent with the data and we can only get a reasonable
fit to the LL for b$\approx$10 km s$^{-1}$.  The L1 system requires a
higher effective b-parameter, suggesting that HI has complex velocity
structure, so that it would not be suitable for a D/H study. 

A further possibility we considered was that neither L1 nor L2 were
responsible for the LL, but that L3, (see Figure 2), was responsible
for the depression.  In order for its LL to cut in at the observed
wavelength, N$_{HI}$ would have to be sufficiently high such that the
high order Lyman lines were strong enough to blend together forming a
substantial drop in the transmitted flux redwards of 912\AA~ in the
rest frame.  However, this would require a high neutral hydrogen
column density which would leave no residual flux at wavelengths
shorter than the LL, which in turn would require the zero level of the
spectrum to be incorrect.  We nevertheless explored this possibility
further by using the EW of L3 to estimate N$_{HI}$ and then comparing
its associated LL with the observed spectrum.  This procedure should
provide an upper limit on N$_{HI}$, since in reality L3 is probably
composed of multiple components.  We found that the N$_{HI}$ derived
in this way did not produce a large enough flux decrement in the
wavelength region shortwards of the observed drop at 2680\AA.

Careful examination of Fig. 2 also suggests that the residual flux in
the data between 2650--2680\AA~ is slightly higher than the synthetic
LL from L2 (or L1) alone.  The spectrum appears to drop to a lower
average flux below $\sim 2640$\AA~ implying the possible presence of
an additional high N$_{HI}$ cloud.

This suggests two steps in the LL, one associated with L2 (z=1.942)
and one with L3 (z=1.870).  We again used VPFIT to model the data with
a 2-component LL system, fixing the two redshifts (z=1.942 and
z=1.870), allowing N$_{HI}$ to vary. The resulting fit can be seen in
panel 4 of Fig. 2.  This procedure results in a good fit to the data,
suggesting that the observed LL is associated with L2 and that there
is indeed a further weak LL absorption associated with L3.

The VPFIT estimate of the L2 column density is
N$_{HI}=(3.1\pm1.0)\times10^{17}$ cm$^{-2}$.  A 2-$\sigma$ upper
limit was also obtained for the N$_{HI}$ of system L3. This was
determined to be N$_{HI}=3.2\times10^{17}$ cm$^{-2}$.  If the LL is
associated with L2 this system may turn out to be suitable for
measuring D/H, because the (tentatively) low effective b-parameter
suggests simple velocity structure and a small velocity dispersion.

Using the two QSO continua described above, we estimated the
parameter representing the flux decrement between the Lyman-$\beta$
and the Lyman-$\alpha$ emission lines, D$_A = 1 - \langle
f_{obs}/f_{cont} \rangle = 0.14\pm0.04$.

\section{Discussion}

We have reported the results of a study of a low/intermediate
resolution spectrum covering the UV/optical/near IR of the HDF-S
QSO J2233$-$606, formed by combining HST/STIS observations with higher
wavelength data from the ANU 2.3m telescope.  The intrinsic QSO
spectral characteristics have been measured from this data.

A study of the absorption properties have revealed multiple CIV
absorption systems, suggesting that either a large structure or
multiple galaxies or clusters intersect the sightline to this QSO.
Structure in the LL absorption is probably caused by the presence of
at least 2 high N(HI) systems, one of which may have properties which
make it suitable for studying D/H.

Despite the low spectral resolution, we have been able to derive
reasonably detailed characteristics of the absorption systems.  Future
high resolution spectra will of course yield far more information.
However, it is already clear from our results that this QSO and the
HDF-S studies of its environs will reveal a host of fascinating new
information about the z$\sim 2$ Universe.

{\section{Acknowledgments}

We particularly thank Dr. B. Boyle for originally suggesting we obtain
the ground-based data. We also thank the director of Mount Stromlo and
Siding Spring Observatories (Australian National University) for the
use of the 2.3m Telescope and Sandra Savaglio for comments.  This work
would not have been possible without the resources allocated to the
HDF-S project by the HST and we thank Dr. H. Ferguson for making the
STIS data available.

\clearpage

\figcaption[final.ps]{Composite spectrum of the HDF-S QSO based on
ground-based ANU 2.3m data (4170 -- 10000\AA) and the HST STIS
spectrum (1600 -- 5700\AA). Also shown is one (the lower) of the two
continua used in the initial determination of N$_{HI}$. \label{fig1}}

\figcaption[hdfs_fig2.ps]{Models of the Lyman limit absorption: the
top panel illustrates the three candidate Lyman-$\alpha$ lines which may
be associated with the LL.  The 2nd and 3rd panels show the LL
region with our models overplotted to determine the velocity structure
of the system associated with the LL.  N$_{HI}$ was calculated as
described in the text, and 3 values of the b-parameter were used
(A=10, B=30 and C=60 km s$^{-1}$) for each of the 2 redshifts.  From
the fits we can see that either L1 (z=1.928) with a high b-parameter
or L2 (z=1.942) with a low b-parameter will fit the data.  The bottom
panel shows the LL region with a model overplotted including the
combined effect at the LL from systems L2 and L3.
\label{fig2}}

\clearpage

\begin{table*}
\begin{center}
\begin{tabular}{lrcc}

\multicolumn{1}{c}{\sf Line}
& \multicolumn{1}{c}{\sf z}
& \multicolumn{1}{c}{\sf EW$_{obs}$ (\AA)}
& \multicolumn{1}{c}{\sf FWHM (km~s$^{-1}$) } \\
\tableline
C IV & 2.236 & 97 & 10370 \\
Al III +  CIII]& 2.238 & 104 & 12560 \\
Mg II& 2.252 & 192 & 9900 \\
all 3& 2.238 &  & \\
& & \\
\end{tabular}
\end{center}

\tablenum{1}
\caption{Emission line parameters measured using the composite ANU
2.3m and HST/STIS spectrum.  The columns give the emission line
identification, the redshift for each line, the observed equivalent width
(EW$_{obs}$) and the velocity width (FWHM).}
\label{tbl1}

\end{table*}

\begin{table*}
\begin{center}
\begin{tabular}{crrrrccr}

\multicolumn{1}{c}{\sf}
& \multicolumn{1}{c}{\sf $\lambda$ (\AA)}
& \multicolumn{1}{c}{\sf $\sigma(\lambda)$}
&\multicolumn{1}{c}{\sf ID}
& \multicolumn{1}{c}{\sf z$_{abs}$}
& \multicolumn{1}{c}{\sf EW$_{obs}$ (\AA)}
& \multicolumn{1}{c}{\sf $\sigma$(EW)}\\
\tableline
& 3388.5 & 1.5 & Ly-$\alpha$ & 1.787 & 1.7 & 0.2 \\
&4316.8 & 1.5 & CIV         & 1.787 & 1.6 & 0.2 \\
&3463.8 & 1.7 & SiIII       & 1.870 & 2.5 & 0.3 \\
&3488.5 & 3.0 & Ly-$\alpha$ & 1.870 & 6.8 & 0.5 \\
&3830.1 & 1.3 & CII         & 1.870 & 1.1 & 0.2 \\
&4444.2 & 1.9 & CIV         & 1.870 & 4.0 & 0.4 \\
&3003.9 & 3.2 & Ly-$\beta$  & 1.928 & 7.3 & 0.5 \\
&3559.2 & 2.0 & Ly-$\alpha$ & 1.928 & 5.1 & 0.5 \\
&3577.3 & 1.6 & Ly-$\alpha$ & 1.942 & 2.2 & 0.3 \\
&4558.7 & 1.8 & CIV         & 1.942 & 3.3 & 0.4 \\
&3741.0 & 1.9 & Ly-$\alpha$ & 2.077 & 4.9 & 0.4 \\
&3895.2 & 1.6 & Ly-$\alpha$ & 2.204 & 2.0 & 0.3 \\
&3973.3 & 0.9 & NV          & 2.204 & 0.9 & 0.2 \\
&4965.3 & 1.7 & CIV         & 2.204 & 2.7 & 0.4 \\
& & \\
\end{tabular}
\end{center}

\tablenum{2}
\caption{Absorption line identifications and parameters, with
1-$\sigma$ errors, for the redshift systems discussed in the text.
EW$_{obs}$ is the observed equivalent width.}
\label{tbl2}

\end{table*}

\clearpage



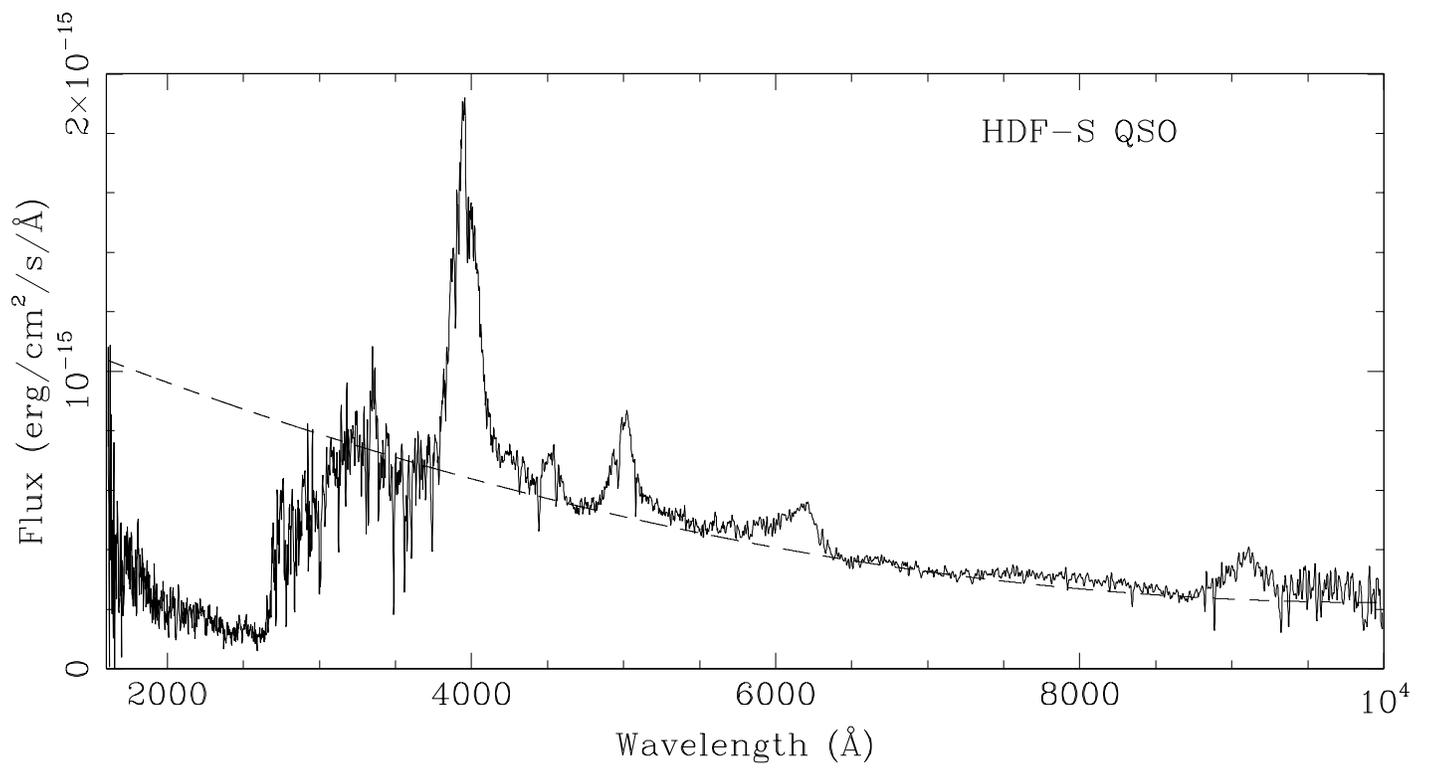
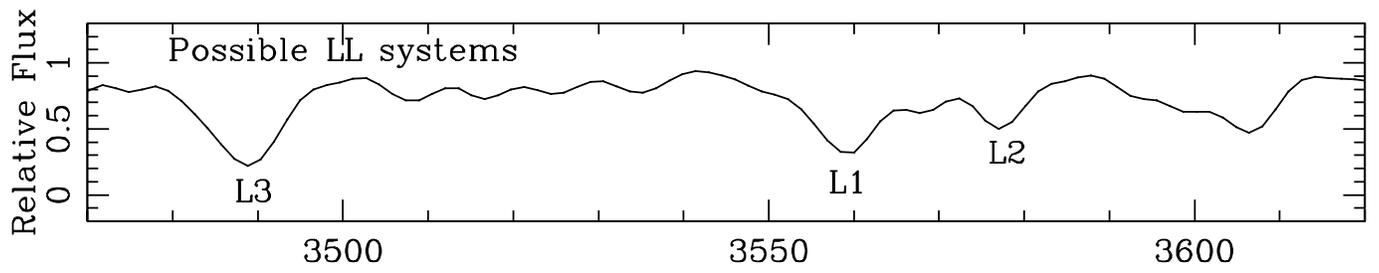
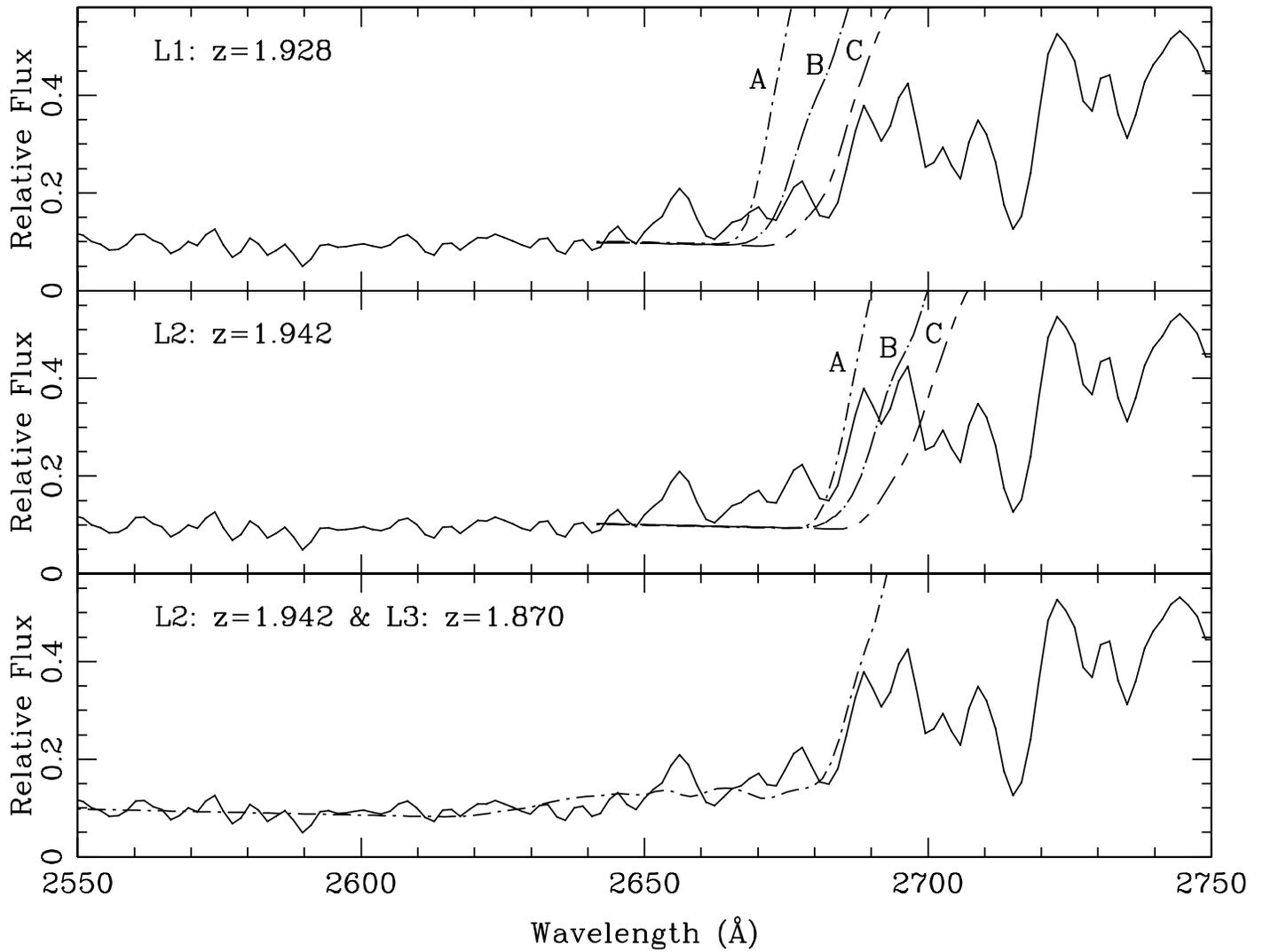
\end{document}